\documentstyle[preprint,aps]{revtex}
\begin{document}
\draft
\title{Quasiparticle Dispersion of the 2D Hubbard Model: \\
{}From an Insulator to a Metal}
\author{R.\ Preuss and W.\ Hanke}
\address{Institut f\"ur Theoretische Physik,
Universit\"at W\"urzburg, Am Hubland\\
D-97074 W\"urzburg, Germany}
\author{W.\ von der Linden}
\address{Max-Planck-Institut f\"ur Plasmaphysik, EURATOM Association\\
D-85740 Garching b.\ M\"unchen, Germany}
\date{\today}
\maketitle
\begin{abstract}
On the basis of Quantum-Monte-Carlo results the evolution of the
spectral weight $A(\vec k, \omega)$ of the two-dimensional Hubbard
model is studied from insulating to metallic behavior. As observed
in recent photoemission experiments for cuprates, the electronic
excitations display essentially doping-independent features:
a quasiparticle-like dispersive narrow band of width of the order of
the exchange interaction $J$ and a broad valence- and conduction-band
background. The continuous evolution is traced back to one and the
same many-body origin: the doping-dependent antiferromagnetic
spin-spin correlation.
\end{abstract}
\par
Recent results of angle resolved photoemission spectroscopy (``ARPES'')
\cite{des93}  revealed strong similarities in the low-energy
excitations of a proto-type insulating copper oxide,
i.e.\ $Sr_2CuO_2Cl_2$ \cite{westb}, and metallic cuprates like
$Bi \ 2212$, $Bi \ 2201$, etc.: in both cases the quasiparticle (QP) band
has rather small dispersion of typically
1 $eV$ width, it is separated from a broad main
valence-band ``background'' (of about 6 $eV$ width)
in much the same way, the $\vec
k$-dispersion is similar and also the intensity modulation as function of
energy is comparable.
Thus, it appears that the dispersive metallic band evolves continuously
from the insulating limit and has a similar physical origin as
the undoped valence band in the cuprates.
This has important consequences for the copper oxides:
the excitation spectrum in the insulating case is decisively determined
by many-body effects, documented by the known difficulties
of one-electron bandstructure calculations  for
the insulating limit \cite{des93}, which then strongly emphasizes a many-body
origin of the QP dispersion also in the metallic case.
\par
In this work Quantum-Monte-Carlo (QMC) results for the angular resolved
photoemission spectral weight $A(\vec k,\omega)$ for the
two-dimensional (2-D) Hubbard model are reported which demonstrate for
this ``generic'' model the above strong similarities between undoped
insulating and doped metallic situations: in particular, we find in both
cases a very similar small dispersive low-energy band, for which the
band width is set by the exchange interaction $J \sim 4t^2/U$ when the
Coulomb correlation $U$ is of order of the non-interacting bandwidth
$8t$.
This new feature is shown in the metallic case to be essentially
unrenormalized (for electronically filled, i.e.\ $\omega<\mu$ states)
from the insulating band, while above the chemical potential $\mu$
additional states with again an energy scale set by $J$ are filled in.
$A(\vec k,\omega)$ is inferred from the QMC data by applying Bayesian
probability theory in the frame of ``quantified maximum entropy''
\cite{gul89,ski90}.
A consistent treatment of hyper-parameters and error-covariances along with
high-quality QMC-data allows to reveal details of the low-lying excitations
which have not been seen before in QMC simulations.
On the basis of these QMC results and the strong similarities to ARPES
data in both insulators and metals, we
argue that the continuous evolution of the QP dispersion
relation observed in the high-${\rm T}_{\rm c}$ cuprates arises from a
common many-body origin, namely from the continuous change of the
spin-spin correlation length $\xi$ and the related changes in the
hole-spin correlations \cite{swz89}.
Similarities to a phenomenology by Kampf and Schrieffer \cite{kas90},
where the magnetic correlation length $\xi$ is an input parameter, are
pointed out.
\par
In the insulating and metallic case, respectively, the QP-like band is
separated similarly from a broad ``incoherent'' background, extending
about $6t$ to $8t$.
The latter corresponds to upper and lower Hubbard bands and follows in
intensity and energy spread roughly the weak-coupling spin-density-wave
(SDW) prediction, as already reported in earlier QMC simulations by
Bulut, Scalapino and White \cite{bsw294,bsw394,bsw94}.
\par
Our results confirm, but also extend these simulation results:
As stated in this previous work the available resolution was
such that when $\vec k$ moved below the Fermi surface only the $6t$ to
$8t$ broad lower Hubbard band could be resolved both for the insulating
\cite{bsw294} and the metallic \cite{bsw394} cases.
However, it was already suspected
there that a narrow QP band exists, since such a
band of width $J$ has been found for one-hole doped in an
AF insulating
t-J model \cite{dnb94,pzs93}.
Narrow QP bands were also observed for finite dopings
in exact-diagonalization calculations of the
one-band Hubbard \cite{dos92,llm92,otk92,mphtb} and t-J \cite{sth91} models,
as well as in QMC-results for the three-band Hubbard model \cite{dwd92}.
\par
The single-band 2-D Hubbard model has the Hamiltonian
\begin{equation}
H = -t \sum_{<i,j>, \sigma} \left(c^\dagger_{i,\sigma} c_{j,\sigma}
+ h.c. \right) + U \sum_i n_{i \uparrow} n_{i \downarrow}
\end{equation}
on a square lattice, where $t$ is the near-neighbor hopping, $c_{i,\sigma}$
destroys an electron of spin $\sigma$ on site $i$ and
$n_{i, \sigma} = c^\dagger_{i,\sigma} c_{i,\sigma}$.
The chemical potential $\mu$ sets the filling $<n>=<n_{i, \uparrow}+
n_{i, \downarrow}>$.
Here we focus on results obtained both for half-filling
$<n>$=1 and other fillings in order to develop a systematic picture
of the low-lying electronic excitations as a function of doping.
The Coulomb correlation $U$ is chosen equal to the bandwidth $8t$ and
as $U=12t$.
\par
In order to obtain from the QMC data for the single-particle
finite-temperature propagator $G(\vec k,\tau)$ the
corresponding spectral weight $A(\vec k,\omega)$ for real frequencies
$\omega$, the following Laplace transform has to be inverted:
\begin{equation}
G(\vec k, \tau)= - \int_{-\infty}^{\infty} A(\vec k,\omega)
\frac{e^{-\tau\omega}}{1+e^{-\beta\omega}} d\omega
\end{equation}
It is by now well established \cite{ssg90,whi91,whi92,lin92,pml94}
that the maximum-entropy method \cite{gul89,ski90}
provides a controlled way to infer the most reliable
$A(\vec k,\omega)$ in the light of the QMC data. To achieve the desired
resolution, it is important to use a likelihood function which takes the
error-covariance matrix of the QMC data and its statistical inaccuracy
consistently into account \cite{lph94}.
The results presented in this letter are based on QMC data with good
statistics, i.e.\ averages over $10^5$ updates of all the
Hubbard-Stratanovich variables result in $G(\vec k,\tau)$'s with
statistical error less than $0.5\%$.
Correlations of the data in imaginary time were considered by making
use of the covariance matrix in the MaxEnt-procedure \cite{gjs91}.
As suggested in previous work by White \cite{whi91}, various moments of the
spectral weight were also incorporated in extracting $A(\vec k,\omega)$.
In order to check on this analytical-continuation procedure detailed
comparisons with $4\times4$ exact diagonalizations \cite{llm92,otk92,mphtb}
have been performed.
\par
The results presented here are for lattices $8\times8$ in size and for
temperatures ranging from $\beta t=3$ $(T=0.33t)$ to $\beta t=10$ $(T=0.1t)$.
Covering this temperature range allows us in effect to study
(at half-filling) a situation where the spin-spin correlation length
$\xi$ is larger (for $\beta t=10$) than the lattice size.
In this case the system behaves as if it were at $T=0$ and develops an
AF gap. For $\beta t=3$, on the other hand, the spin-spin
correlation length is shorter than our finite lattice and, consequently,
the gap is diminished and metallic fluctuations exist \cite{whi92}.
As we will see, the latter situation is especially useful in
interpreting the QMC data in the metallic (doped) regime and to relate
them to those of the insulating (half-filled) case.
\par
We start in Fig.\ 1 by examining the single-particle spectral weight for
$U=8t$, $\beta t=10$, at half-filling, i.e.\ $<n>=1$. Fig.\ 1(a) gives
a 3-D plot of $A(\vec k,\omega)$ versus $\omega$ for $\vec k$-values out of
the Brillouin zone, whereas Fig.\ 1(b) summarizes these results in the
usual ``bandstructure'' $\omega$ versus $\vec k$ plot. Here dark (white)
areas correspond to a large (small) spectral weight.
\par
We observe in all spectra (also in the $U=12t$, half-filled case, presented
in Fig.\ 3) two general features:
One is that $A(\vec k,\omega)$ contains a rather dispersion-less
``incoherent background'', extending both for electronically occupied
($\omega < \mu$) states and unoccupied ($\omega > \mu$) states over $\sim 6t$
($\sim 6 \ eV$) in the $U=8t$ case. The new structure, which was not
previously resolved in QMC work
is a dispersing structure at low energies with small width of the
order of $J$, which defines the gap
$\Delta$ and which, (at least for $U=12t$), is well separated from the
higher energy background.
\par
The splitting in the low-energy ``band'' and the higher-energy
``background'' is especially pronounced near $\Gamma$-(for $\omega<\mu$) and
M-(for $\omega>\mu$) points due to a relative weight
shift from negative to positive energies as $\vec k$ moves through
X or equally through ($\pi/2,\pi/2$), the midpoint between $\Gamma$ and M.
The overall weight distribution in $A(\vec k,\omega)$ follows roughly the SDW
prediction as found in the QMC calculation by Bulut et al.\ \cite{bsw294}:
the total integrated weight in the SDW approximation
$<n_{\vec k}^{SDW}>=\int_{-\infty}^0A^{SDW}(\vec k,\omega) d\omega$, is
in good accord with the QMC momentum distribution \cite{bsw294}.
\par
However, the dispersion of the structure near the gap does not follow the SDW
prediction: its dispersion has a significant (about a factor of 2 for
$U=8t$) smaller width set by the value of $J$. This result is in good accord
with the dispersion and width found for the low energy ``foot'' in recent
angle-resolved photoemission (ARPES) data \cite{westb} and t-J model results
(there $t \approx 0.4 \ eV$) \cite{dnb94}.
This is illustrated in Fig.\ 1(b), where the low-energy peaks in
$A(\vec k,\omega)$ are fitted by (full and dotted lines)
$E_{\vec k}=\Delta + J/2 (cos k_x + cos k_y)^2$,
with $\Delta=2.4t$, rather than by the SDW (strong-coupling) result, i.e.\
$E_{\vec k} = \sqrt{\epsilon_{k}^{2}+\Delta^2}\cong\Delta+J(cos k_x +
cos k_y)^2$.
The overall agreement between the ARPES width and the Hubbard model data
(for $t\approx 1 eV$)
is significant because it shows in fact that the energy scale of
the low-lying insulating band is controlled by many-body effects
beyond the mean-field SDW result.
\par
Our findings for the spectral-weight dispersion $\omega(\vec k)$ are
schematically summarized in Fig.\ 2(a), which emphasizes again the
different energy scales (J, U, SDW\cite{comment}) involved.
These energy scales become particularly pronounced in the
($U=12t$)-case in Fig.\ 3.
\par
Before moving to the doped situation, we note that the $\beta t=3$,
$U=12t$ result in Fig.\ 3 does have the valence-band maximum at the
M-point and not at the X- or ($\pi/2,\pi/2$)-points in contrast to
the $\beta t=10$ results for both $U=8t$ and $U=12t$ (not shown).
This at first puzzling
``high-temperature'' result reflects the fact that at $\beta t=3$ the
spin-spin correlation length $\xi$ is about a factor of 2.5 smaller than
the QMC lattice extension.
The system then shows precursor effects of a metal, which move the
spectral weight (valence-band) maximum -- in agreement with the metallic
situation in Fig.\ 4 -- to the M-point. Otherwise, the low-energy band is
found to be essentially unaffected by changing temperature from $T=0.1t$
to $T=0.33t$.
\par
Keeping this in mind, we consider in Fig.\ 4 the low-energy electronic
structure in the metallic regime for doping $<n>$=0.95.
4(a) shows again the 3-D plot of $A(\vec k,\omega)$ and 4(b)
the dispersion relation with the degree of shading representing
the intensity of $A(\vec k,\omega)$, as in Fig.\ 1(b).
Like in the half-filled case, we observe two general features, which
are both seen in recent photoemission experiments \cite{des93}: a
broad ``background'' of $\approx 4t-6t$ spanning the lower and upper
Hubbard band and a pronounced low-energy ``foot'' of significantly
smaller width, which is clearly resolved between $\Gamma$$\rightarrow$X
and  $\Gamma$$\rightarrow$$(\pi/2,\pi/2)$.
The situation is depicted schematically in Fig.2b.
Width and dispersion of the low-energy ``band'' are also in good accord
with exact diagonalization results of $4\times4$ clusters
\cite{llm92,mphtb}.
\par
The results for the doped case have several important implications:
First, they
reveal that the lowest energy ``band'' in the insulator and the ``band'' that
crosses the chemical potential, $\omega=\mu$, in the
hole-doped metal are rather similar: the low-energy band is separated
from the broad valence-band background (``LHB'' (lower Hubbard band) in the
schematic drawings of Fig.\ 2) in the same way; it
has similar dispersion and it has a similar intensity modulation as a
function of energy.
Thus, the lowest-energy metallic band appears to be effected by similar
many-body physics as in the insulating regime, namely magnetic
correlations connected with the (now) short-range AF order.
This is not in contradiction but instead substantiated by the fact that
the metallic band develops its maximum at the M-point: as pointed out
above, this happens as soon as the spin-spin correlation length $\xi$ is
smaller than the length ($L$=8), a situation obtained for the filling
$<n>$=0.95.
\par
Furthermore, our QMC data display similarities to results derived by Kampf and
Schrieffer \cite{kas90}.
There a phenomenological model for the spin susceptibility was
used to extract the leading-order contribution to the self-energy from
AF-fluctuations.
The spin-spin correlation length $\xi$ was an input parameter
to the model and by varying this correlation length the model evolved
continuously from the insulating AF to the Fermi-liquid regime.
For decreasing correlation length weight was transferred into the QP
peak from incoherent backgrounds, which (for $\xi$ large)
developed into the upper and lower Hubbard bands (our structures ``UHB'' and
``LHB'').
This is schematically summarized in Fig.\ 2(b).
\par
Another noteworthy feature of the doped, metallic situation is that the
intensity change (but not the width) as a function of
$\vec k$ for the higher-energy background in $A(\vec k,\omega)$ still
follows essentially the AF SDW
picture with, in particular ``shadow bands'' resulting from the AF
short-range order being clearly visible at $\Gamma$-($\omega-\mu\sim-5t$) and
M-($\omega-\mu\sim+7t$) points.
Even remnants can be detected of folded back
``shadow bands'' near the M-point for $\omega<\mu$, which in the SDW-picture
have much less oscillator strength and spectral intensity than the
original band (between $\Gamma$$\rightarrow$X) \cite{hmd94}.
The findings confirm again to a certain extent the phenomenological work
by Kampf and Schrieffer \cite{kas90}.
Finally, we would like to mention that an important detail of the QMC
data, the rather extended flatness of the energy band near the
X($\pi,0$)-point, which is in good agreement with ARPES experiments of
Dessau et al. \cite{des93} for $Bi \ 2212$, has previously been resolved
in QMC work both for the one-band \cite{bsw394} and the three-band
\cite{dwd92,ppm94} Hubbard models.
This rather extended
flatness, extending like the ARPES data not only into
X$\rightarrow$$\Gamma$, but also into the X$\rightarrow$M regions (as
displayed in Fig.\ 3) is already inherent in the undoped low-energy
structure near X, a fact which has recently also been found in 2-D t-J
model studies by Dagotto et al.\ \cite{dnb94}.
Its extension, in particular into X$\rightarrow$M
direction, is not consistent with available one-electron band
calculations \cite{des93}.
It can be explained by a conventional self-energy diagrammatic
analysis summing over the leading spin-fluctuation diagrams \cite{ppm94}.
It is thus a many-body effect related to magnetic correlations
consistent with the arguments given in this work.
\par
In summary, we have studied the evolution of the 2-D Hubbard model from
insulator to metal in terms of the  electronic spectral weight, obtained
from the maximum-entropy analytic continuation of QMC data.
These results, combined with recent ARPES data, can be taken as strong
indication that the QP dispersion of the high-${\rm T}_{\rm c}$
compounds, not only in the insulating limit but -- particularly -- in
the metallic situation, has a many-body origin:
the coupling of the quasiparticles to antiferromagnetic correlations.
\par
We would like to thank R.\ Laughlin, S.\ Maekawa, A.\ Muramatsu,
D.\ Poilblanc, H.\ Schulz, Z.-X.\ Shen and, particularly, D.J.\ Scalapino
for instructive discussions.
The calculations were performed at the HLRZ J\"ulich and at the
LRZ M\"unchen.

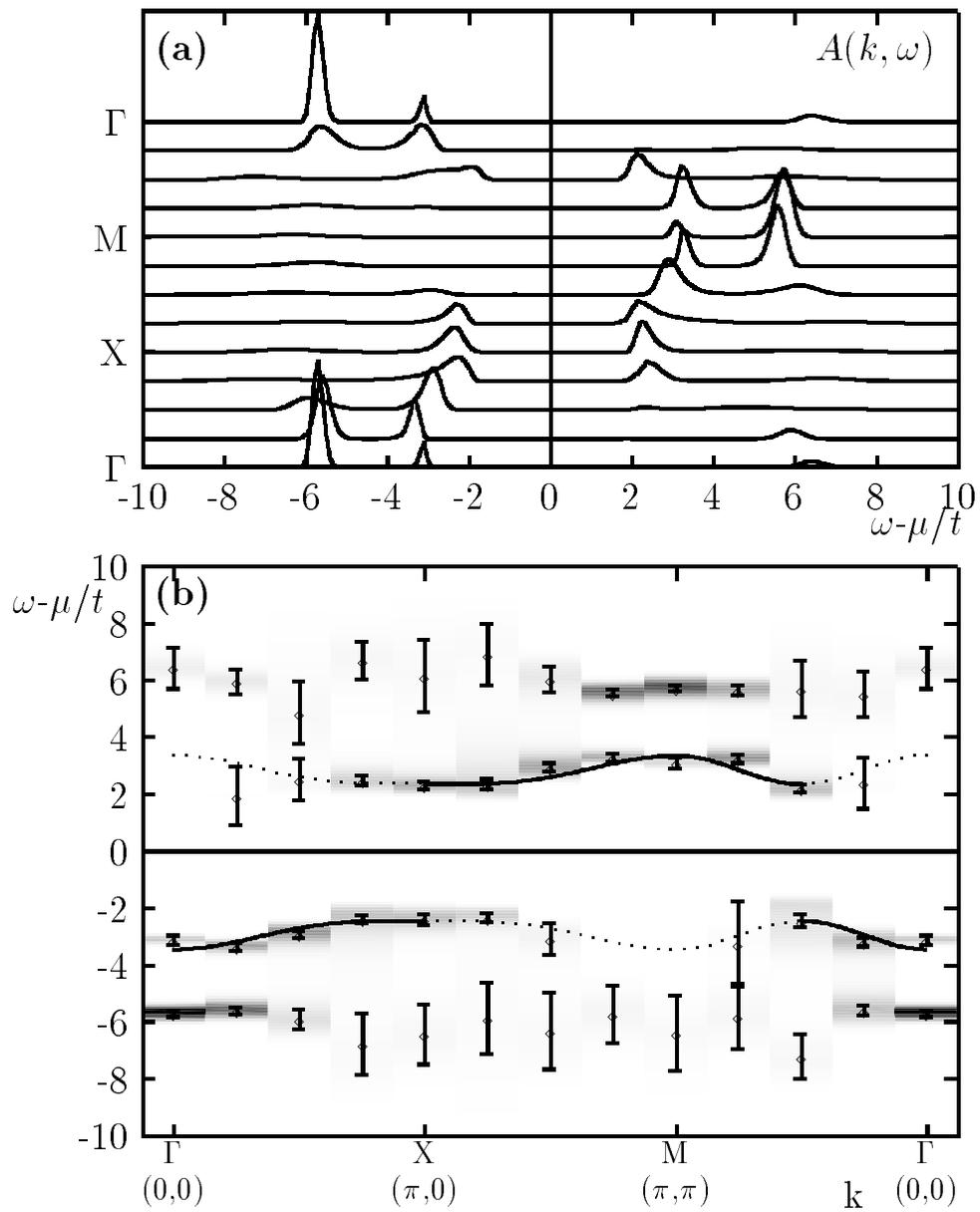
\begin{figure}
\caption{
Single-particle excitation for the $8\times8$ Hubbard model at
half-filling, $U=8t$,
$\beta t=10$:
(a) The single particle spectral weight $A(\vec k,\omega)$ versus $\omega$
and $\vec k$;
(b) $\omega$ versus $\vec k$ ``bandstructure'', where sizable structure
in $A(\vec k,\omega)$ is represented by strongly shaded areas and peaks by
error bars.
}
\end{figure}

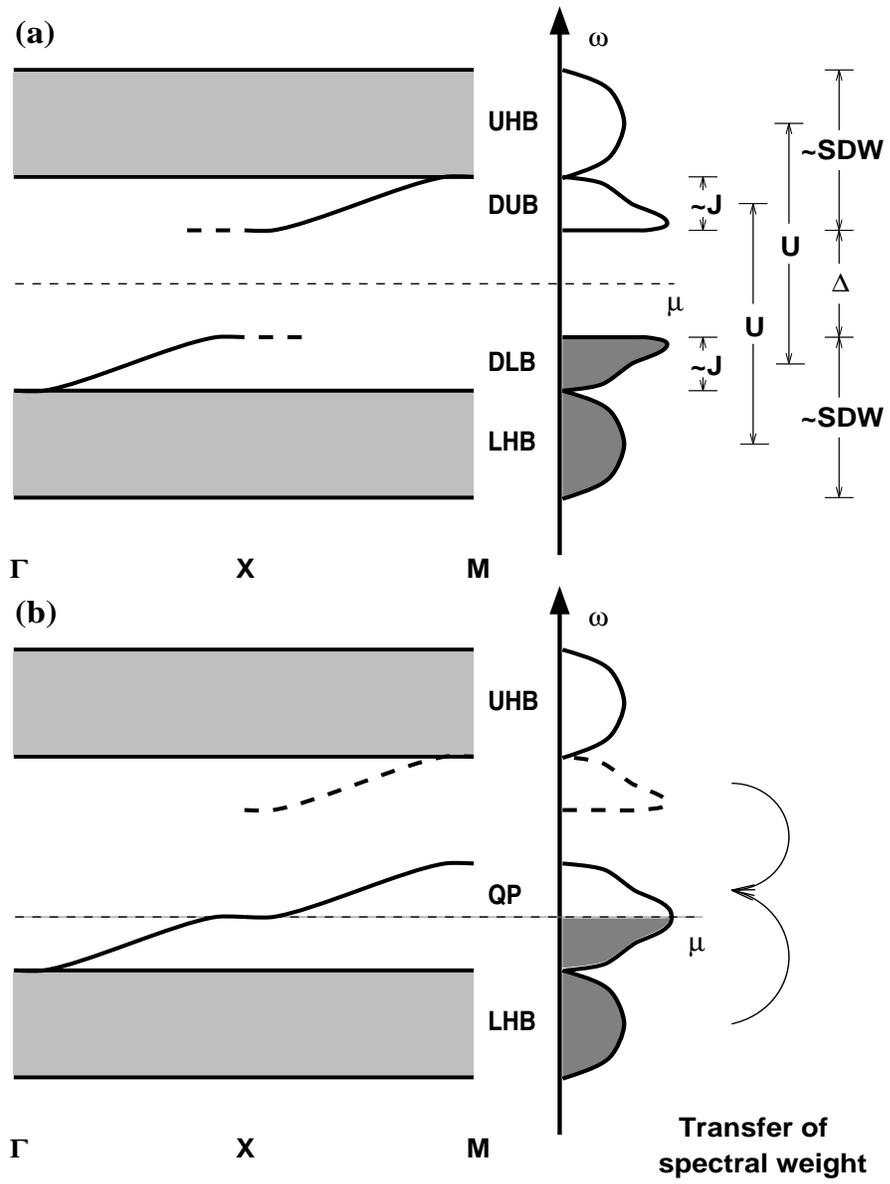
\begin{figure}
\caption{
Schematic plot of the bandstructure:
(a) the insulator and the different energy scales (see text) involved;
(b) for the metallic situation.
}
\end{figure}

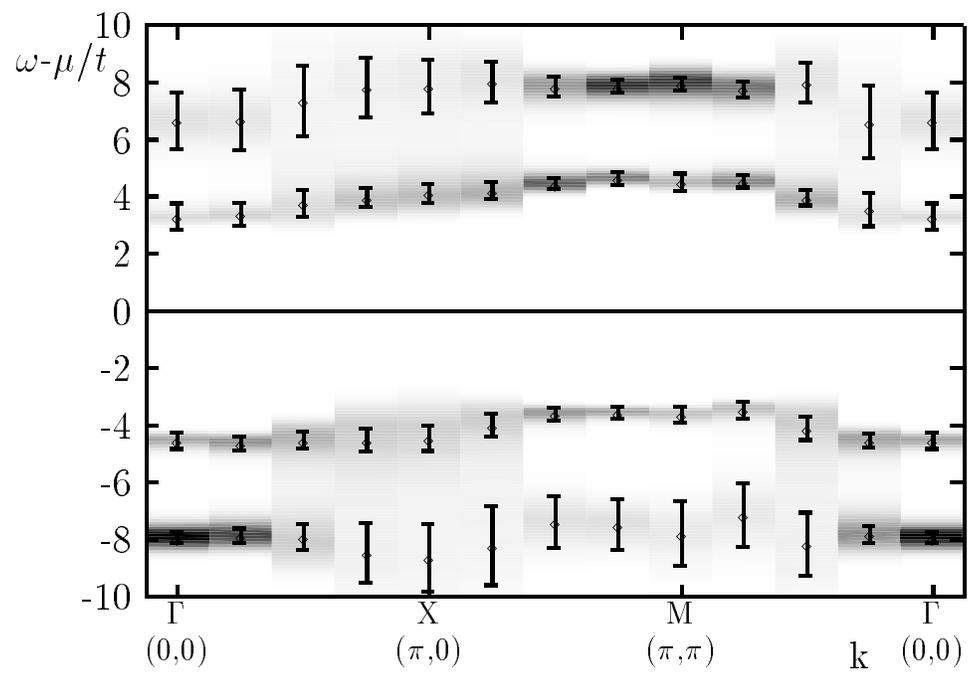
\begin{figure}
\caption{
Single ``bandstructure'' $\omega$ versus $\vec k$ for $U=12t$,
$\beta t=3$. Again strongly shaded areas correspond to maxima of
$A(\vec k,\omega)$.
}
\end{figure}

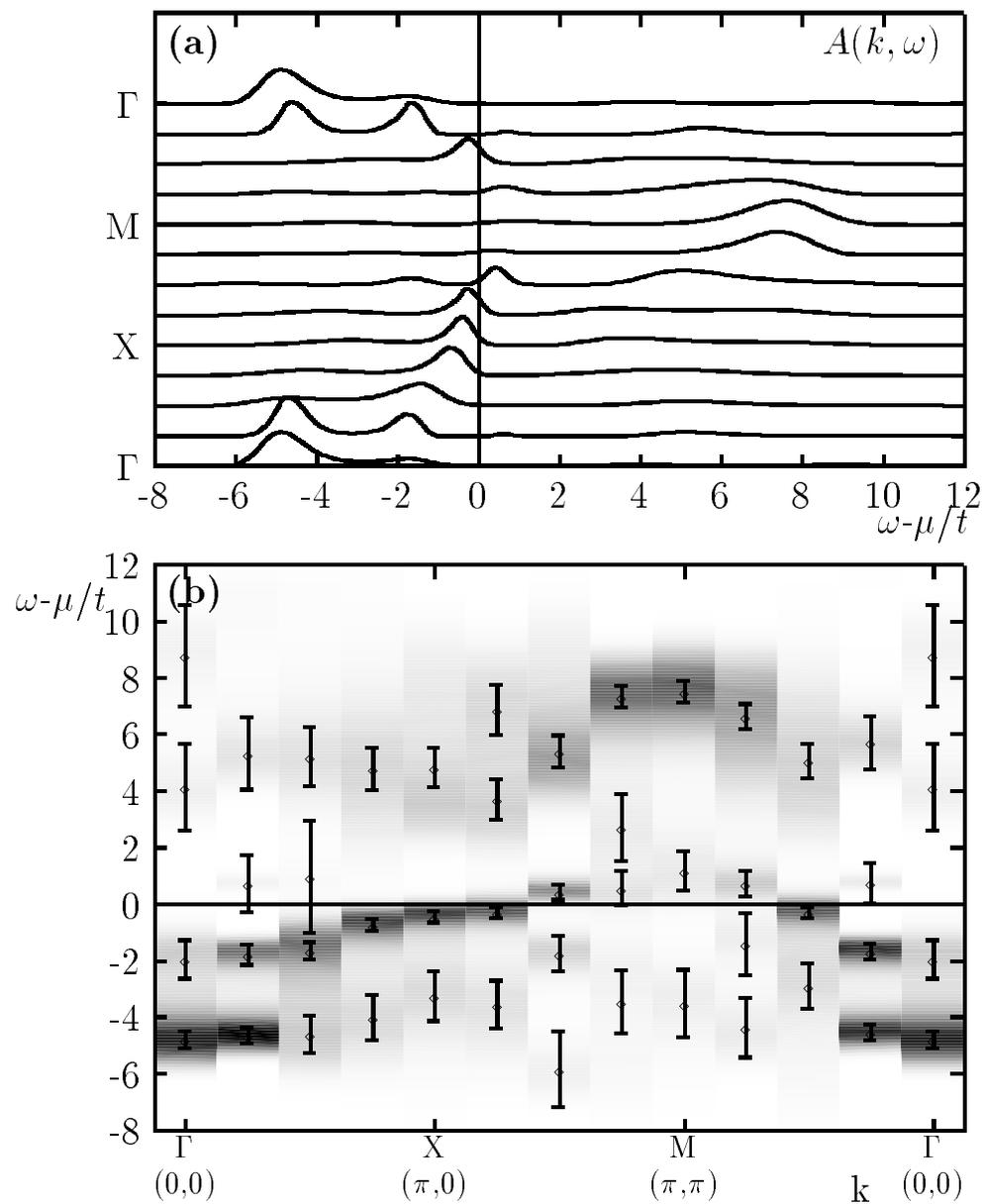
\begin{figure}
\caption{
The same as Fig.\ 1, but now for doping $<n>$=0.95 and
$\beta t=3$, $U=8t$:
(a) $A(\vec k,\omega)$;
(b) $\omega$ versus $\vec k$ ``bandstructure''.
}
\end{figure}


\begin{references}
\bibitem{comment} ``SDW'' stands here for the total integrated weight,
                not for the energy spread $\sim 6t$ which is
                significantly larger than the SDW band width.
\end{references}
\end{document}